# Is Green Enough? A Remote Sensing Assessment of Environmental Impacts and Green Commitments at Beijing Daxing International Airport

Haorui Wang, Bo Zhao
University of Washington, Seattle

ABSTRACT

Beijing Daxing International Airport has been promoted as a model of green infrastructure under China's ecological modernization agenda. Featuring energy-efficient design, renewable energy systems, and smart environmental controls, the airport embodies multiple green commitments. This study evaluates its environmental outcomes using multi-source remote sensing data—including NDVI, NDBI, Land Surface Temperature (LST), VIIRS night-time lights, and PM2.5—from 2014 to 2019. Through spatial and temporal comparisons, we assess landscape-level changes during and after construction. Findings indicate partial gains from green initiatives but also reveal substantial vegetation loss, increased built-up surfaces, and intensified surface temperatures. The results suggest a gap between sustainable design and ecological impact. We propose a remote-sensing-based framework for evaluating future infrastructure projects, emphasizing the need for spatially explicit, independent monitoring to ensure environmental accountability.

## 1. Introduction

As large-scale infrastructure projects continue to proliferate globally, the concept of "green development" has gained significant traction (Li et al., 2019). Airports, in particular, have emerged as both celebrated symbols of connectivity and productivity, and as sites of intense scrutiny for their ecological costs. In recent years, governments and developers have increasingly framed new airport projects as "green" by incorporating renewable energy systems, energy-efficient architecture, and environmental mitigation strategies. These initiatives reflect a growing institutional commitment to sustainable urbanization, but they also raise critical questions about what "greenness" truly entails—and how it should be evaluated. Beijing Daxing International Airport, inaugurated in 2019, stands as one of China's most prominent examples of green-branded infrastructure. Celebrated for its futuristic design, intelligent energy systems, and LEED-compliant construction, the airport was positioned as a flagship project under the national vision of "eco-civilization" (Zhang & Fu, 2023). Its official sustainability profile highlights features such as large-scale solar installations, rainwater harvesting systems, natural lighting strategies, and landscape greening. However, despite these efforts, it remains unclear whether such design and planning choices translate into meaningful ecological benefits at the regional level. The gap between process-based greenness—the deployment of sustainable technologies—and outcome-based greenness—measurable improvements in ecological conditions—has yet to be fully explored.

      Conventional environmental impact assessments, though widely applied in the infrastructure sector, often rely on project-based reporting and offer only a static snapshot of ecological status. These limitations pose challenges when attempting to assess the broader, cumulative, and dynamic effects of megaprojects on surrounding landscapes. In contrast, remote sensing offers a spatially expansive and temporally continuous approach to environmental monitoring. Indicators such as the Normalized Difference Vegetation Index (NDVI), the Normalized Difference Built-up Index (NDBI), Land Surface Temperature (LST), and VIIRS night-time light intensity provide meaningful proxies for assessing landscape change, construction intensity, and human activity. Therefore, this paper uses a multi-indicator remote sensing framework to evaluate the environmental impacts of Daxing Airport from

2014 to 2019. Focusing on the airport's construction period, it analyzes whether and how surface-level ecological conditions changed during and after development. By comparing satellite-based indicators before and after the airport's completion, the study interrogates the effectiveness of green design principles in mitigating ecological disturbance. More broadly, it contributes to ongoing discussions about how green infrastructure should be defined, implemented, and measured in practice. Through this case, we argue for the necessity of outcome-based evaluation frameworks that extend beyond symbolic sustainability narratives and are grounded in empirical, spatially resolved data.

## 2. Literature review

Environmental considerations have been central to airport planning and development since the late 1960s, supported by a range of regulatory frameworks that aim to minimize negative ecological impacts (Culberson, 2011). In the United States, foundational policies such as the National Environmental Policy Act (NEPA, 1969) and the Clean Air Act (1970) require comprehensive environmental assessments for major infrastructure projects, including airports. Similarly, the European Union mandates Environmental Impact Assessments (EIA) and Strategic Environmental Assessments (SEA) as part of its project approval process (Culberson, 2011). These regulations cover multiple dimensions, such as air quality, biodiversity, noise pollution, waste management, and broader social and economic concerns. In recent years, beyond compliance, many airport projects have begun to incorporate broader sustainability goals—aiming to reduce their carbon footprints, improve energy and water efficiency, and implement more effective waste management strategies (Raimundo et al., 2023).

One of the most immediate environmental impacts of airports concerns air quality. Airports contribute a wide array of pollutants into the atmosphere—including benzene, carbon monoxide (CO), nitrogen dioxide ($NO_2$), ozone ($O_3$), lead (Pb), sulfur dioxide ($SO_2$), and particulate matter (PM)—through aircraft operations, auxiliary power units, ground support vehicles, and ongoing construction activities (Culberson, 2011). These emissions not only degrade air quality but also pose direct risks to human health. Moreover, the aviation industry contributes substantially to global greenhouse gas emissions, making airport operations a notable factor in climate change (Raimundo et al., 2023). In addition to atmospheric concerns, the construction and operation of airports often disrupt local ecosystems. Biodiversity impacts are common, as natural habitats are destroyed or fragmented, particularly in areas previously used as farmland or wetlands. Noise pollution—another frequent byproduct of airport operations—has well-documented health effects. It can lead to sleep disruption, impaired communication, learning difficulties in children, heightened stress, and even cardiovascular problems (Culberson, 2011). Long-term noise exposure is especially problematic in densely populated or residential areas surrounding major airports. Equally important are the hydrological and land-use changes caused by airport development. Paving over large areas of land with asphalt and concrete interrupts natural drainage and creates impermeable surfaces that exacerbate flooding risks. In response, the "sponge airport" concept has been developed to manage stormwater more sustainably. This green infrastructure model incorporates permeable pavements, rainwater harvesting systems, green roofs, and vegetated swales to increase the absorption and reuse of stormwater. Beijing Daxing International Airport was one of the first to apply this model on a large scale. Modeling by Peng et al. (2021) shows that these low-impact development (LID) features can significantly reduce surface runoff and improve local drainage performance.

While Daxing Airport was promoted as a cutting-edge example of green infrastructure at its opening in 2019, independent studies using remote sensing data have begun to critically examine its actual environmental outcomes. Wan et al. (2022), for instance, analyzed Landsat-8 imagery and found that land surface temperatures in the airport zone increased substantially following construction. This urban heat island (UHI) effect extended up to 7.5 km beyond the airport and was attributed to expanded impervious surfaces, reduced vegetation, and heightened human activity. Similarly, Yang et al. (2022) applied a remote sensing ecological index to compare conditions before and after construction and observed a clear decline in ecological quality around the airport, largely driven by vegetation loss and landscape hardening. These findings suggest a disconnect between design-based green intentions and outcome-based ecological results. This pattern is not unique to Beijing. Research from airports worldwide has documented similar trends. In tropical cities like Singapore, for example, Jusuf et al. (2007) found that the extensive paved surfaces of Changi Airport contributed to some of the highest recorded daytime surface temperatures, just behind those of industrial zones. Other studies in Dalian, China and Osaka, Japan similarly confirm that airports act as significant heat sources, with their impact shaped by location and surrounding land use (Yu et al., 2018). In Europe, De'Donato et al. (2007) compared urban and suburban airports in Italian cities and found that proximity to urban centers intensified the heat island effect, emphasizing the need for strategic site selection. To mitigate these environmental pressures—particularly UHI effects—planners and researchers have proposed several interventions. Lai et al. (2019), for example, reviewed urban cooling strategies and found that increasing vegetation cover and using reflective building materials can significantly lower ambient temperatures. Incorporating such features into airport design may not only reduce heat but also enhance overall sustainability. These strategies complement ongoing efforts to manage emissions and improve resource efficiency, demonstrating that environmental impacts can be partially mitigated through thoughtful design.

While airports undoubtedly support economic growth and regional connectivity, they also produce a wide range of ecological impacts, including degraded air and water quality, greenhouse gas emissions, biodiversity disruption, and urban heating. Addressing these issues requires a combination of regulatory frameworks, green design innovations, and rigorous post-construction evaluation. Tools like remote sensing are particularly well-suited for this purpose, offering an independent, spatially explicit, and scalable approach to monitoring environmental change over time. As airports continue to expand globally, especially in rapidly developing regions, ensuring that "green" promises translate into measurable outcomes will be essential for achieving truly sustainable infrastructure.

## 3 Data and Methods

Beijing Daxing International Airport and its surrounding region serve as the primary study area for this research. The airport, completed in 2019, is situated between Beijing's Daxing District and Langfang in Hebei Province. As a prominent case of large-scale, state-led infrastructure framed by green development promises, this site presents a unique opportunity to evaluate the environmental impacts of construction and operation through spatially explicit, time-sensitive remote sensing analysis. To capture the effects of land cover change, vegetation degradation, and urban thermal expansion, we analyzed satellite imagery over a region centered on the airport with a buffer wide enough to include surrounding development zones.

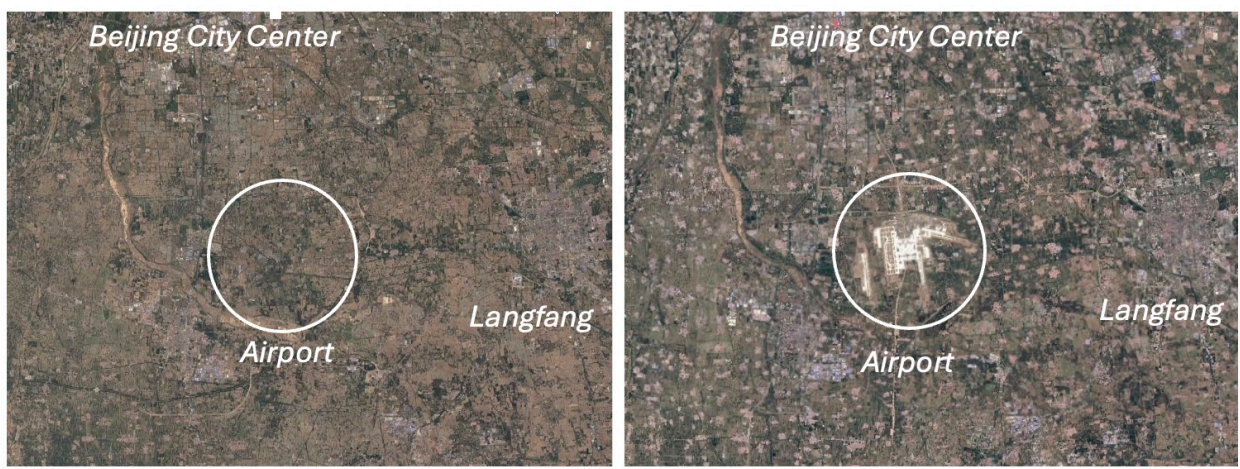

**Figure 1** Comparison of Natural Color Satellite Images of the study area between 2014 Pre-Construction (left) and 2017 Post-Construction (right).

This study draws on multiple publicly available satellite and geospatial datasets, summarized in Table 1. To assess environmental impacts and the performance of green commitments, we extracted a set of key indicators from the satellite imagery. These include vegetation cover, built-up area expansion, surface temperature, night-time activity intensity, and air pollution levels. Specifically, we used the Normalized Difference Vegetation Index (NDVI) to evaluate vegetation change and greening efforts, and the Normalized Difference Built-up Index (NDBI) to monitor the spatial expansion of constructed surfaces. Land Surface Temperature (LST) was derived to assess urban heat island effects, while VIIRS night-time lights were used as a proxy for urbanization intensity and human activity. PM2.5 concentrations were retrieved to reflect potential changes in air quality during and after the construction period. The calculation of these indicators followed standardized remote sensing procedures, including radiometric correction, index transformation, and pixel-based analysis.

A crucial part of the data processing involved ensuring the temporal and spatial reliability of the imagery. Since satellite observations are often hindered by cloud contamination, especially during certain seasons, we applied a cloud filtering strategy to exclude low-quality scenes. Additionally, when imagery for a target date was unavailable or unusable due to cloud cover, we adopted a temporal backtracking approach. This method involved selecting the nearest available image in time, ideally within the same season, to maintain consistency across the dataset. These strategies helped mitigate common remote sensing limitations and ensured the integrity of our time-series comparison. In addition, the temporal design of this study focused on two critical moments: the pre-construction stage in 2014 and the post-construction phase in 2019. These years mark the baseline and immediate aftermath of the airport's development and thus serve as reference points for evaluating environmental transformation. The choice of these years enables a controlled comparison of land surface conditions before and after major anthropogenic interventions associated with the airport project.

**Table 1.** Overview of Remote Sensing and Geospatial Data Sources Used in the Study

| Data Source | Resolution | Use in Study |
|:---:|:---:|:---|
| Landsat 8 | 30 m | NDVI, NDBI– building, vegetation, and heat index tracking |
| VIIRS | ~500 m | Night-time lights – proxy for urban activity intensity |
| MODIS | 1 km | LST – high-frequency land surface temperature tracking |

To analyze and communicate the results, we employed a variety of visualization and statistical methods. Spatial patterns were illustrated using comparative maps, allowing for direct visual interpretation of indicator changes between the two periods. To complement these maps, we used histograms to examine the distribution of indicator values across the study area and calculated mean differences to quantify the magnitude of change. Where available, time-series visualizations were also used to trace dynamic trends in environmental indicators across multiple dates. These techniques not only provided clarity in presenting the findings but also offered critical insights into the effectiveness—and limitations—of Daxing Airport's green infrastructure claims.

## 4. Results

This section presents the spatial and temporal patterns of key environmental indicators derived from multi-source remote sensing data spanning the period from 2014 to 2019. Through the integration of visual analysis and statistical interpretation, we assess how the construction and subsequent operation of Beijing Daxing International Airport have influenced land surface conditions across both the core airport footprint and its surrounding buffer zones.

As shown in the left column of Figure 2, Land Surface Temperature (LST) increased significantly in both spatial distribution and intensity, with the most dramatic changes concentrated within the airport's central infrastructure zone. The 2019 thermal imagery highlights elevated surface temperatures along the main runways, terminal aprons, and adjacent hardscaped areas, with warmer gradients radiating outward into neighboring regions. In contrast, the 2014 imagery shows a cooler thermal profile, with most pixel values remaining below 310K. By 2019, contiguous zones exceeding 313K had emerged, particularly within newly paved and built-up areas. This shift illustrates a classic Urban Heat Island (UHI) effect, most likely resulting from widespread surface sealing, increased structural density, and reduced vegetative cooling through evapotranspiration.

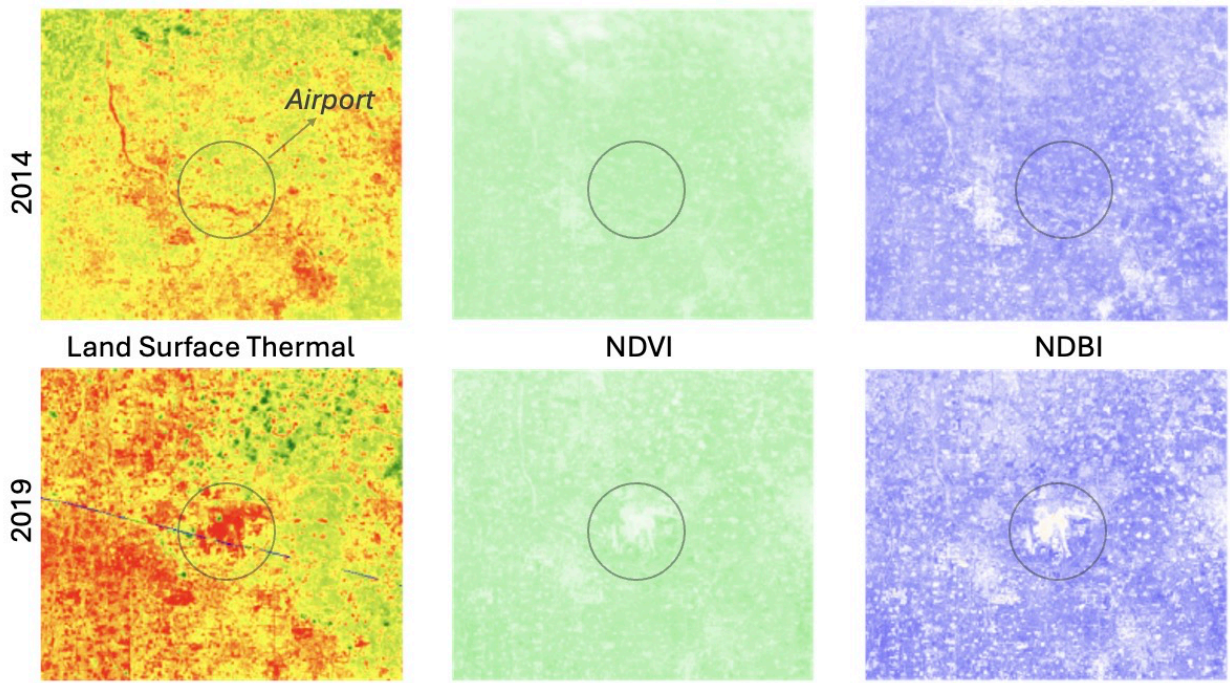

**Figure 2.** Spatial comparison of environmental indicators in the Daxing Airport region between 2014 (top row) and 2019 (bottom row). From left to right: Land Surface Temperature (LST), Normalized Difference Vegetation Index (NDVI), and VIIRS night-time light intensity. The 2019 maps show increased surface temperatures, reduced vegetation, and brighter night-time illumination, indicating intensified land use and urbanization following airport construction.

Normalized Difference Vegetation Index (NDVI) analysis reveals a substantial loss of vegetation during the early stages of airport construction, followed by a modest degree of re-greening. As shown in the middle column of Figure 2, the 2014 imagery is dominated by NDVI values above 0.35, indicating widespread agricultural or semi-natural vegetation cover. By 2019, these high-value zones had contracted significantly, particularly within the core airport footprint and along major access corridors. In their place, NDVI values closer to 0.15 or lower emerged, clearly reflecting vegetation clearance and land conversion. This spatial shift is especially evident along the eastern and southeastern flanks, where vegetation loss is most pronounced. Temporal NDVI values, visualized in Figure 3, further quantify this pattern: following a sharp drop in 2015 that coincides with site clearing, a brief recovery is observed in 2016–2017, but the index stabilizes afterward. By 2019, the average NDVI remains approximately 12% lower than its 2014 baseline, suggesting that ecological restoration has been limited or largely cosmetic.

In parallel with vegetation decline, the Normalized Difference Built-up Index (NDBI) shows a clear increase in built-up intensity. As indicated in the right column of Figure 2, 2014 NDBI values exceeding 0.2 appear only in scattered patches along major roads. By 2019, however, the central airport area displays expansive contiguous zones with NDBI values above 0.4. These built-up signals extend outward to the north and west, reflecting new industrial and logistical development surrounding the airport. The combined pattern of high NDBI and low NDVI, as shown in Figure 3, confirms a large-scale and largely irreversible transformation of greenfield land into hard urban infrastructure.

The environmental consequences of large-scale land-use conversion include long-term, and often irreversible, degradation of soil and water quality—resulting from compaction, erosion, and potential contamination during the construction process. These impacts unfold within the ecological context of Daxing District, which hosts a forest coverage rate of 33.80% and a green space ratio of 46.05%. The area supports a diverse range of flora and fauna, including 125 plant species, 128 tree species, 29 animal species, and 211 bird species (The People's Government of Daxing District, 2023). The observed NDVI decline from approximately 0.300 to below 0.150 represents a serious and potentially irreversible threat to this biodiversity, despite ongoing efforts at restoration through landscaping or replanting. In addition to vegetation loss, biodiversity degradation has been intensified by increased noise pollution and the airport's wildlife deterrent measures. Elevated levels of human activity—such as jet engine operations, construction noise, and traffic—disturb local species, altering habitats and triggering forced migration or behavioral stress. To reduce the risk of bird strikes, the airport has implemented laser deterrents and electronic systems that emit simulated eagle calls (Awen, 2019). While these measures improve aviation safety, they have also driven native bird populations out of the area, further reducing the region's ecological richness.

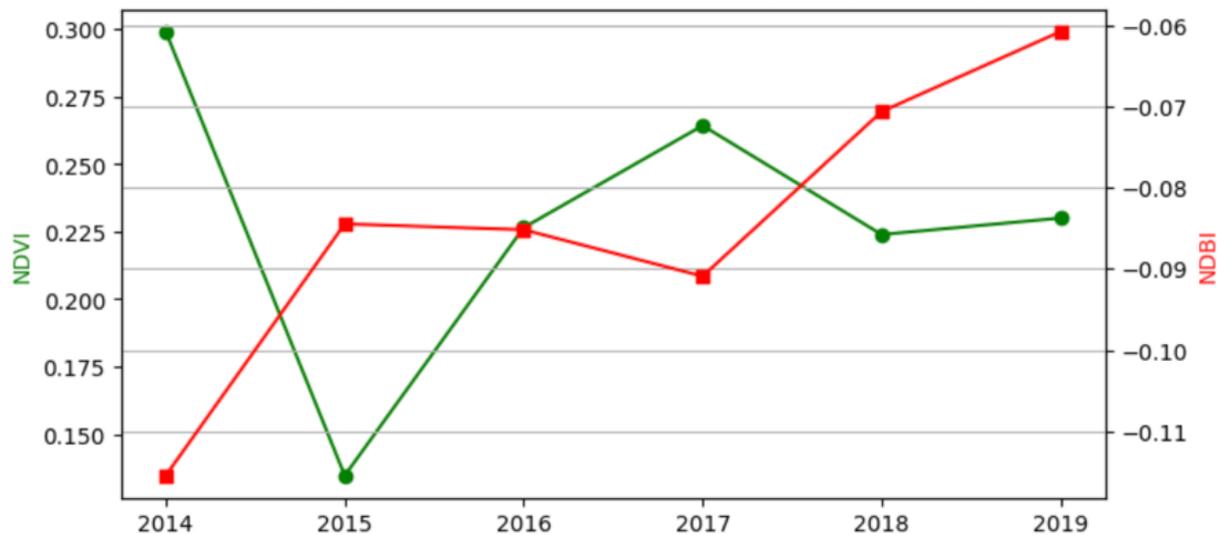

*Figure 3. Times series of NDVI and NDBI values from 2014 to 2019*

Beyond the direct transformation caused by the airport itself, surrounding areas have also undergone rapid economic development. This growth has led to further expansion of settlements, the construction of artificial green spaces such as parks, and improvements in habitat and field management. These interventions have contributed to a modest recovery in NDVI values, reflecting the replanting of vegetation and landscape greening. However, this phenomenon reflects a familiar "build-first, restore-later" pattern in urban development. Economic growth acts as a double-edged sword: while it brings environmental stress through construction, energy consumption, and land-use change, it also generates the financial capacity and political will to invest in beautification and sustainability efforts. In the case of Daxing, the area's growth exemplifies a typical developmental arc—initial ecological degradation followed by compensatory greening. Yet, as ecological balance is shaped by long-term natural processes, such short-term remedies often fail to fully restore biodiversity, soil quality, and habitat integrity. Though partial recovery may look promising on paper, much of the original ecological complexity is lost. These trade-offs were seldom acknowledged during early policy decision-making, when the emphasis was placed on economic potential, national prestige, and "green" branding statistics, rather than long-term ecological sustainability.

While LST, NDVI, and NDBI highlight the direct environmental impacts of airport construction, the broader effects of regional economic growth are more clearly visualized through satellite observations of night-time light. As shown in Figure 4, the 2019 VIIRS night light imagery reveals substantial growth compared to 2014, particularly in four areas: (1) the airport zone itself, (2) the northern parts of Daxing District, (3) residential clusters southwest of the airport, and (4) Langfang City to the east. Night-time light serves as a proxy for economic activity and regional GDP (Pérez-Sindín et al., 2021). While some increase reflects natural population-driven urbanization, the scale and distribution of illumination suggest that the airport acted as a key catalyst for the spatial intensification of development. As illustrated in Figure 5, night-time light levels rose gradually during the airport's construction period but spiked dramatically between 2018 and 2019, coinciding with its operational launch. This surge underscores the airport's role as a powerful economic stimulus. However, with such growth comes environmental cost. Increased land surface temperatures and urban heat effects are just one part of a broader set of degradations. Many other consequences—such as water stress, waste accumulation, and habitat fragmentation—are less easily detected by satellite and remain underexamined. Short-term economic gains, as suggested by the Environmental Kuznets Curve (EKC), often correlate with higher levels of carbon emissions (Kahuthu, 2006), worsening air quality (Rao & Yan, 2020), and increased solid waste generation (Kaza et al., 2018). These findings emphasize that growth-driven development must be carefully balanced with long-term ecological stewardship.

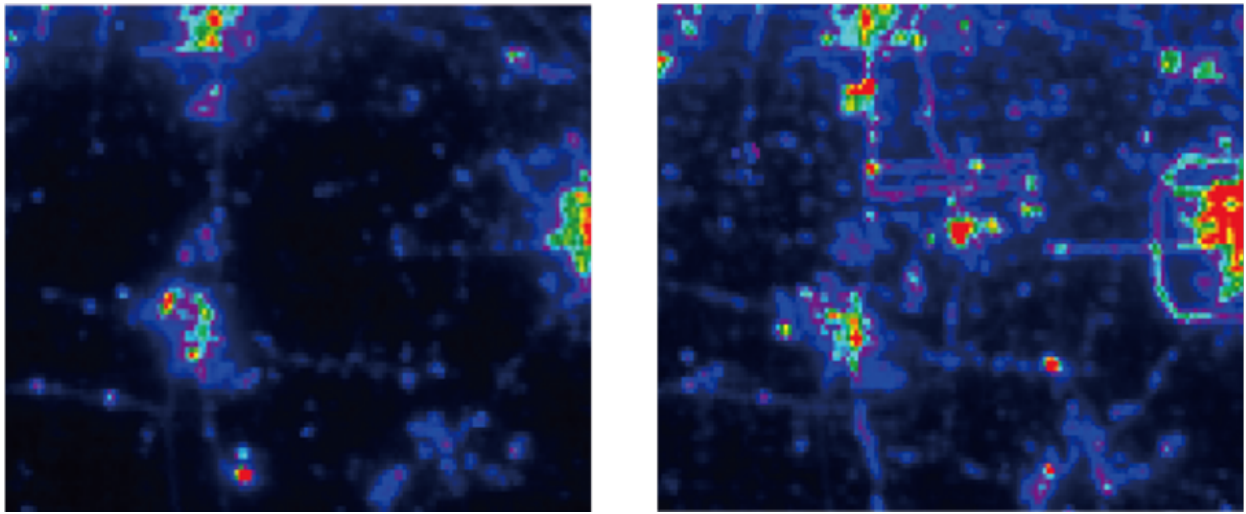

**Figure 4.** *VIIRS night-time light comparison (2014 vs 2019), showing spatial growth of light intensity in both core and periphery development zones.*

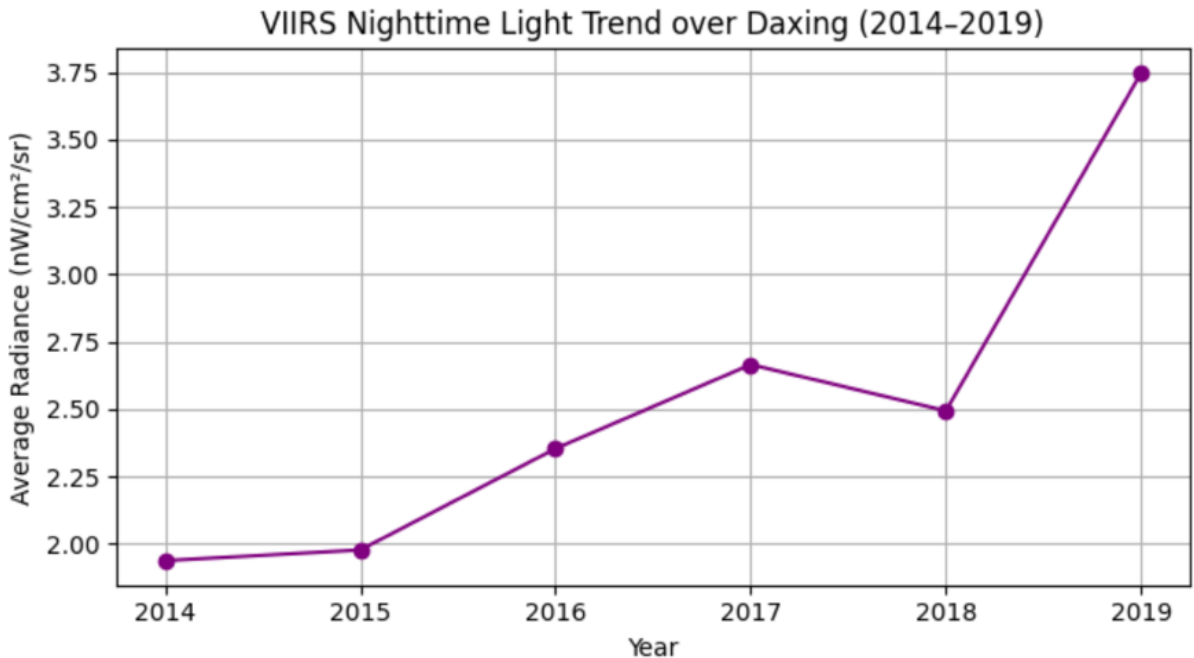

**Figure 5.** *Times series of VIIRS Nighttime Light values from 2014 to 2019*

      While satellite-derived PM2.5 estimates were included in the analysis, the observed changes between 2014 and 2019 were relatively modest. Across most of the airport region, concentrations remained in the 30–60 µg/m³ range during both years, with only slight increases detected near highway intersections and cargo-handling zones. Due to the limited spatial resolution and contrast of the atmospheric datasets, PM2.5 patterns were not sufficiently distinct for visual mapping and are therefore excluded from figure-based presentations. Instead, they are summarized in the comparative indicator table below.

      This summary table aggregates the average values across all key environmental indicators. LST, NDVI, and NDBI each show consistent directional changes, reinforcing the spatial trends visualized earlier. Night-time light intensity also exhibits a substantial increase, indicative of accelerated urban and economic activity. In contrast, PM2.5 variations were minor. Taken together, these results confirm that the construction and operation of Beijing Daxing International Airport have driven wide-ranging environmental transformations—across thermal, ecological, structural, and socio-economic dimensions.

**Table 2.** Summary of indicator changes between 2014 and 2019 in the Daxing Airport region.

| Indicator | 2014 Value (mean) | 2019 Value (mean) | Interpretation |
|---|---|---|---|
| **LST** | ~310K | ~313K | Significant surface warming |
| **NDVI** | ~0.300 | ~0.263 | Vegetation loss and partial greening |
| **NDBI** | ~0.100 | ~0.135 | Built-up surface expansion |
| **Night Lights** | low | high | Intensified night-time activity and infrastructure |

| Indicator | 2014 Value (mean) | 2019 Value (mean) | Interpretation |
|---|---|---|---|
| PM2.5 | ~40 µg/m³ | ~45 µg/m³ | Modest air quality degradation |

The most prominent shift among all indicators is the ~3K increase in median Land Surface Temperature (LST) between 2014 and 2019, confirming the emergence of Urban Heat Islands (UHI) centered around the airport. This localized thermal intensification reflects a pattern commonly observed in airport-adjacent urban development zones (Yang et al., 2016; Cleland et al., 2023). The conversion of vegetated surfaces to impervious materials, combined with increased anthropogenic heat emissions, appears to be the primary driver of this transformation.

NDVI values declined sharply during the airport's initial site-clearing phase in 2014–2015. Although some degree of vegetative recovery followed, the 2019 levels remained significantly lower than the pre-construction baseline. This trend suggests that while landscaping and replanting may have been carried out—perhaps to meet aesthetic or regulatory expectations—true ecological recovery in terms of habitat complexity, biodiversity, and carbon storage remained incomplete. The region's ecological richness diminished, further impacted by airport wildlife control measures such as sound cannons and laser deterrents aimed at reducing bird strikes (Daxing District Government, 2023).

NDBI results further validate the transformation of land use, with built-up expansion concentrated in the airport core and extending into peripheral zones. Areas with high NDBI values overlap spatially with regions showing NDVI decline and LST increase, pointing to a cumulative and systemic shift in surface characteristics. Similar patterns have been observed in other large-scale airport developments in East Asia, where infrastructure growth triggers substantial ecological and thermal impacts (Jusuf et al., 2007).

Finally, the surge in night-time light intensity between 2014 and 2019 reflects broader economic activation. The airport itself became a major source of illumination, while neighboring towns—including Langfang and northern Daxing—also brightened significantly. This aligns with increased infrastructure usage and regional GDP expansion, consistent with state-led urban development agendas. However, this growth also signals heightened energy demand and rising levels of light pollution, adding another layer of environmental pressure (Pérez-Sindín et al., 2021).

## 5 Discussion

This study reveals the multilayered environmental consequences of constructing Beijing Daxing International Airport. While the airport has been promoted as a green infrastructure exemplar through its use of solar panels, rainwater recycling systems, and other energy-saving features, remote sensing indicators show substantial landscape transformations that raise important questions about how "green" such megaprojects truly are.

By analyzing changes in land surface temperature (LST), vegetation cover (NDVI), built-up area (NDBI), and night-time lights, we distinguish between process-based greenness—the environmental practices employed during planning and construction—and outcome-based greenness, which reflects actual ecological results. Though efforts were made to plant vegetation and manage environmental impact, the results suggest that ecological degradation outpaced restoration. Specifically, the exacerbation of the UHI effect and vegetation loss uncovers that habitat destruction leads to permanent damages to the environment. In addition, night light imagery illustrates the regional development from a macro-perspective, corroborating the intensification of human activities and economic growth in the study area. While the economy may flourish, the indirect environmental impacts such as increased energy consumption, greenhouse gas emissions, light pollution, and resource consumption must also be considered. As shown, airports are agents of economic growth but also drivers of environmental change. This echoes arguments from infrastructure economics that megaprojects function as regional growth engines with high environmental stakes (SEO Amsterdam Economics 2024), and environmental degradation tends to rise during early phases of development before potentially stabilizing at higher income levels (Kahuthu 2006) —a pattern consistent with our findings.

Remote sensing proves particularly valuable in assessing environmental outcomes independently of project reports or self-declared sustainability credentials. By applying consistent spatial and temporal metrics, we evaluated whether green claims resulted in green outcomes. This external, data-driven audit approach reinforces the potential for satellite analysis to serve as a feedback mechanism for large-scale infrastructure evaluation. Further research could be done with better spatial, temporal, and spectral resolution, such as delving deeper in air pollution with Aerosol Optical Thickness (AOT). It is suggested that more output-based greenness assessments are made for other airports and large-scale infrastructure projects and observed from a longer time interval to investigate long-term effects. Furthermore, future studies could benefit from integrating remote sensing with other methodologies that are more ground-based for precise micro-level data. A limitation of this study is that the time interval used, 2014-2019, cannot be extended longer to, say, 2025, because the COVID-19 pandemic was a major disruptor that confounds the data after 2019, rendering analysis difficult to lead to any conclusive results. Overall, our findings support the case for a more robust green infrastructure evaluation framework in China and elsewhere. We propose that future airport and transport projects incorporate both process-based and outcome-based indicators, with remote sensing serving as a long-term monitoring tool. Without clear, independently measurable benchmarks, even well-intentioned green infrastructure projects risk falling short of their ecological promises.

## 6. Conclusion

This study has assessed the environmental impacts of Beijing Daxing International Airport by analyzing multi-temporal satellite imagery across five indicators: land surface temperature (LST), vegetation index (NDVI), built-up index (NDBI), night-time light intensity, and estimated PM2.5 concentrations. The analysis reveals substantial surface-level changes that occurred between 2014 and 2019, particularly during and after the airport's construction.

The findings demonstrate that while the airport has been officially framed as a model of green infrastructure, the actual environmental outcomes are more complex. A pronounced urban heat island emerged in the airport core, vegetation cover was significantly reduced with only partial recovery, built-up intensity increased visibly, and night-time light patterns expanded spatially and in intensity. These signals of environmental transformation suggest that process-oriented sustainability measures—such as the incorporation of energy-efficient design features—did not fully translate into positive landscape-level ecological outcomes. Importantly, this research distinguishes between green intent and green effect, highlighting the gap between sustainability narratives and actual ecological performance. It shows that infrastructural greening does not necessarily prevent broader forms of ecological disruption, especially when land transformation occurs at a regional scale. Remote sensing proves to be a valuable tool in critically assessing such large-scale projects. Its capacity to independently verify environmental trends over time and space offers an important complement to conventional environmental impact assessments, which may be limited by narrow temporal scopes or institutional bias. This method also supports policy learning by providing a baseline for long-term ecological monitoring.

Looking forward, the case of Daxing suggests the need for more outcome-based accountability frameworks in green infrastructure evaluation. Rather than relying solely on design intentions or self-reported sustainability metrics, urban megaprojects should be evaluated using consistent, transparent, and spatially resolved indicators of environmental change. As China and other countries continue to invest in massive transportation infrastructure, this kind of evidence-based approach will be essential to ensuring that green development is not just symbolic—but substantive.